\documentclass[aps,pre,english,twocolumn,showpacs,preprintnumbers,amsmath,amssymb,floatfix,superscriptaddress]{revtex4-1}
\usepackage[T1]{fontenc}
\usepackage[latin9]{inputenc}
\usepackage{graphicx}
\usepackage{amssymb}
\usepackage{babel}
\makeatletter

\usepackage[version=3]{mhchem}

\usepackage{color}

\usepackage{ulem}

\makeatother
\usepackage{babel}
\makeatother

\begin{document}

\title{Intermolecular Casimir-Polder Forces in Water and near Surfaces}

\author{Priyadarshini Thiyam}
\email{thiyam@kth.se}
\affiliation{Department of Materials Science and Engineering, Royal Institute of Technology, SE-100 44 Stockholm, Sweden}

\author{Clas Persson}
\affiliation{Department of Materials Science and Engineering, Royal Institute of Technology, SE-100 44 Stockholm, Sweden}
\affiliation{Department of Physics, University of Oslo, P.O. Box 1048 Blindern, NO-0316 Oslo, Norway}
\affiliation{Centre for Materials Science and Nanotechnology, University of Oslo, P.O. Box 1048 Blindern, NO-0316 Oslo, Norway}

\author{Bo E. Sernelius}
\affiliation{Division of Theory and Modeling, Department of Physics, Chemistry and Biology, Link{\"o}ping University, SE-581 83 Link{\"o}ping, Sweden}

\author{Drew F. Parsons}
\affiliation{Department of Applied Mathematics, Australian National University, Canberra, Australia}

\author{Anders Malthe-S{\o}renssen}
\affiliation{Department of Physics, University of Oslo, P.O. Box 1048 Blindern, NO-0316 Oslo, Norway}

\author{Mathias Bostr{\"o}m}
\email{Mathias.Bostrom@smn.uio.no}
\affiliation{Centre for Materials Science and Nanotechnology, University of Oslo, P.O. Box 1048 Blindern, NO-0316 Oslo, Norway}
\affiliation{Department of Materials Science and Engineering, Royal Institute of Technology, SE-100 44 Stockholm, Sweden}
\affiliation{Department of Energy and Process Engineering, Norwegian University of Science and Technology, NO-7491 Trondheim, Norway}

\begin{abstract}
The Casimir-Polder force is an important long range interaction involved in adsorption and desorption of molecules in fluids. We explore Casimir-Polder interactions between methane molecules in water, and between a molecule in water near SiO$_2$ and hexane surfaces. Inclusion of the finite molecular size in the expression for the Casimir-Polder energy leads to estimates of the dispersion contribution to the binding energies between molecules and between one molecule and a planar surface.
\end{abstract}

\pacs{34.20.Cf; 42.50.Lc; 03.70.+k; 88.20.fq}

\maketitle



\section{Introduction}

Methane gas extracted from shale gas systems is emerging as an important source of energy with a low carbon footprint, and shale gas may serve as an important transient energy source in a global transition to a low emission economy. Shale gas extraction may pose evironmental challenges -- for example, it has been suggested that methane may seep to surrounding aquifers during hydraulic fracturing and horizontal drilling \,\cite{Osborn, Vidic} -- but shale gas systems may also provide new opportunities to address environmental challenges. In shale gas systems methane is found in fractures, dissolved in fluids 
or kerogen, adsorbed on mineral and kerogen surfaces and stored in nanoporous spaces within the kerogen. It has been suggested that CO$_2$ injection may be used to enhance gas production because CO$_2$ binds more strongly to relevant surfaces than methane, but our understanding of the related fundamental processes is still lacking. One important contibution to adsorption forces comes from the van der Waals and Casimir-Polder interactions. It is therefore important to study and understand how methane interacts with different substances and how methane behaves near surfaces in water, in order to develop efficient methods for enhanced hydrocarbon production and simultaneous CO$_2$ storage in shale-gas systems.

In  a liquid it is well known that van der Waals interactions between polarizable particles or surfaces may be either attractive or repulsive.\,\cite{Maha,Ninhb,Buhmann12a,Buhmann12b}  
 In this work we explore the Casimir-Polder interaction between methane molecules in water as well as the  interaction of a methane molecule near different surfaces solved in water, for example SiO$_2$, hexane, air, {\it etc}. Intermolecular dispersion interactions between two polarizable particles in water that account for finite size have in the past only been considered within an approximate series expanded theory.\,\cite{Maha} We will demonstrate that the van der Waals contribution to the binding energy of two molecules in water is very similar in expanded and non-expanded theories.  
We find here that methane molecules in water attract to each other, and a methane molecule in water is attracted to SiO$_2$ surface while it experiences repulsion near a hexane surface. 
The new non-expanded theory presented for molecule-molecule dispersion interaction shows that previous work that used  series expansion for the van der Waals contribution to the binding energy of two molecules were rather accurate. This motivates us to use the expanded theory for molecule-surface van der Waals interaction. The theory of Casimir-Polder interaction between finite size polarizable molecules is given in Sec. II. We will furthermore investigate the dispersion contribution to the binding energy of methane molecules near different surfaces solved in water. The known theory of molecule-surface interaction in presence of a background medium is briefly described in Sec. III. In Sec. IV we describe the calculation of the excess polarizability of methane molecules in water. The correct way to calculate excess polarizability that is consistent with the Green's functions follows from the book by Parsegian.\,\cite{Pars}   We discuss the dielectric functions for SiO$_2$, water, and hexane in  Sec. V. We present our main results in Sec. VI and end with some final conclusions. Details about the Green's function elements and some discussions about the multipolar dispersion forces are presented in the two appendices.

\section{Long range Casimir-Polder Energy Between Finite Size Polarizable Particles in Water}

The  Casimir-Polder interaction between two polarizable particles in  a liquid is\,\cite{Safari,Maha,Mahanty3,Sernelius,Mitchell,MahNin}

\begin{equation}
U(\rho ) = {k_B}T  \sum\limits_{n = 0}^{\infty}{'} \ln|\tilde 1- \tilde \alpha_1^* (i \xi _n)  \tilde T_1   \tilde \alpha_2^* (i \xi _n) \tilde T_2  | 
\label{Eq1}
\end{equation}
where $\tilde 1$ is the identity matrix.

For two equal isotropic particles with finite size one obtains the following Casimir-Polder interaction energy\,\cite{Sernelius}
\begin{equation}
U(\rho ) = {k_B}T \sum_{j=x,y,z} \sum\limits_{n = 0}^{\infty}{'} \ln[1- \alpha_i^* (i{\xi _n)}^2    T_{jj}^2],
\label{Eq2}
\end{equation}
where $\alpha_i^*(i\xi_n )$ is the excess  polarisability of particle $i$ at the Matsubara frequencies  $\xi_n=2 \pi k_B T n/\hbar$.\,\cite{Dzya,Sernelius}. We define $k_B$ as the Boltzmann constant, $T$  the temperature, and the prime indicates that the $n=0$ term shall be divided by 2. When this expression is series expanded one obtains the textbook result presented for example in the book by Parsegian\,\cite{Pars} 
\begin{equation}
U(\rho ) = -{k_B}T \sum_{j=x,y,z} \sum\limits_{n = 0}^{\infty}{'} \alpha_i^* (i{\xi _n)}^2    T_{jj}^2,
\label{Eq3}
\end{equation}
where only the first term in the logarithmic expansion of Eq. (2) is considered. In the non-expanded theory, however, the full expression is taken into account numerically. We consider, as an interesting case, a Gaussian function to represent the finite spread of the polarization cloud of real atom or molecule. We have derived, following the formalism developed by Mahanty and Ninham\,\cite{Maha,Mahanty3} (see Appendix A) the Green's function elements ($T_{jj}$) that account for retardation, background media, and finite size

\begin{equation}
\begin{array}{*{20}{l}}
{{T_{xx}}\left( {\rho |i\xi_n } \right) = {T_{yy}}\left( {\rho |i\xi_n } \right)}\\
{ =  - \frac{{\exp \left[ {{{\left( {\frac{\xi_n }{c}} \right)}^2}{{\left( {\frac{a}{2}} \right)}^2}} \right]}}{{2\rho  }}\left\{ {\left[ {{{\left( {\frac{\xi_n }{c}} \right)}^2} + \left( {\frac{\xi_n }{c}} \right)\frac{1}{\rho } + {{\left( {\frac{1}{\rho }} \right)}^2}} \right]} \right.}\\
{\begin{array}{*{20}{l}}
{ \times \left[ {1 - {\rm{erf}}\left( {\frac{\xi_n }{c}\frac{a}{2} - \frac{\rho }{a}} \right)} \right]\exp \left( { - \frac{\xi_n }{c}\rho } \right)}\\
{ - \left[ {{{\left( {\frac{\xi_n }{c}} \right)}^2} - \left( {\frac{\xi_n }{c}} \right)\frac{1}{\rho } + {{\left( {\frac{1}{\rho }} \right)}^2}} \right]}
\end{array}}\\
{\begin{array}{*{20}{l}}
{ \times \left[ {1 - {\rm{erf}}\left( {\frac{\xi_n }{c}\frac{a}{2} + \frac{\rho }{a}} \right)} \right]\exp \left( {\frac{\xi_n }{c}\rho } \right)}\\
{\left. { - \frac{4}{{a\rho \sqrt \pi  }}\exp \left[ { - {{\left( {\frac{\xi_n }{c}} \right)}^2}{{\left( {\frac{a}{2}} \right)}^2} - {{\left( {\frac{\rho }{a}} \right)}^2}} \right]} \right\},}
\end{array}}
\end{array}
\label{Eq4}
\end{equation}
\begin{equation}
\begin{array}{*{20}{l}}
{{T_{zz}}\left( {\rho |i\xi_n } \right)}\\
{\begin{array}{*{20}{l}}
{ = \frac{{\exp \left[ {{{\left( {\frac{\xi_n }{c}} \right)}^2}{{\left( {\frac{a}{2}} \right)}^2}} \right]}}{  \rho }\left\{ {\left[ {\left( {\frac{\xi_n }{c}} \right)\frac{1}{\rho } + {{\left( {\frac{1}{\rho }} \right)}^2}} \right]} \right.}\\
{ \times \left[ {1 - {\rm{erf}}\left( {\frac{\xi_n }{c}\frac{a}{2} - \frac{\rho }{a}} \right)} \right]\exp \left( { - \frac{\xi_n }{c}\rho } \right)}
\end{array}}\\
{ + \left[ {\left( {\frac{\xi_n }{c}} \right)\frac{1}{\rho } - {{\left( {\frac{1}{\rho }} \right)}^2}} \right]\left[ {1 - {\rm{erf}}\left( {\frac{\xi_n }{c}\frac{a}{2} + \frac{\rho }{a}} \right)} \right]\exp \left( {\frac{\xi_n }{c}\rho } \right)}\\
{ - \left. {\frac{4}{{a\rho \sqrt \pi  }}\left( {1 + {{\left( {\frac{\rho }{a}} \right)}^2}} \right)\exp \left[ { - {{\left( {\frac{\xi_n }{c}} \right)}^2}{{\left( {\frac{a}{2}} \right)}^2} - {{\left( {\frac{\rho }{a}} \right)}^2}} \right]} \right\}.}
\end{array}
\label{Eq5}
\end{equation}
Here {$a=1.505 $\AA} is the Gaussian radius of the methane molecule, $c=c_0/\sqrt{\varepsilon(\xi_n)}$; $c_0$ is the velocity of light in vacuum and $\varepsilon(i \xi_n)$ is the dielectric function of water for imaginary frequencies.  The excess polarizability consistent with this definition of the Green's function elements is given in Sec. V. Except when the molecules come so close that their electron clouds start to overlap one can ignore finite size effects, {\it i.e.} $a \to \ 0$ and then finds\,\cite{MahNin}

\begin{equation}
\begin{array}{*{20}{l}}
{T_{xx}^n(i{\xi_n}) = T_{yy}^n(i{\xi _n}) =  - (\frac{{\xi_n^2}}{{{c^2}}} + \frac{{{\xi_n}}}{{\rho c}} + \frac{1}{{{\rho ^2}}})\frac{{{e^{ - {\xi_n}\rho /c}}}}{{\rho } 
},}\\
{T_{zz}^n(i{\xi _n}) = 2(\frac{1}{{{\rho ^2}}} + \frac{{{\xi_n}}}{{\rho c}})\frac{{{e^{ - {\xi_n}\rho /c}}}}{{\rho }}.}
\end{array}
\label{Eq6}
\end{equation}
This diagonal form is obtained if the $z$-axis is defined to point along the line joining the two particles. We will onwards in this work use both the theory with the size effects and the corresponding one without the size effects. 

\section{Non-retarded van der Waals Energy of a Molecule near an Interface}

We also consider the non-retarded van der Waals energy between a finite size methane molecule and an interface. The expression for this is given in the work of Ninham and co-workers.\,\cite{Maha,ParsPCCP,BostNin} The dispersion interaction free energy of a molecule in water at a distance $\rho$ from an interface between water and a second medium with dielectric function $\varepsilon_{surface}$ is

\begin{equation}
\begin{array}{l}
U = \frac{{Bf(\rho )}}{{{\rho ^3}}},\\
B = \frac{{{k_B}T}}{2}\sum\limits_{n = 0}^\infty  {'} {{\alpha ^*}(i{\xi _n})\frac{{{\varepsilon _{water}} - {\varepsilon _{surface}}}}{{{\varepsilon _{water}} + {\varepsilon _{surface}}}}} ,\\
f(\rho ) = 1 + \frac{{2\rho }}{{a\sqrt \pi  }}(\frac{{2{\rho ^2}}}{{{a^2}}} - 1)\exp ( - \frac{{{\rho ^2}}}{{{a^2}}}) - (1 + \frac{{4{\rho ^4}}}{{{a^4}}}){\rm{erfc}}(\frac{\rho }{{\rm{a}}}).
\end{array}
\label{Eq7}
\end{equation}
Including the effects of finite size in the formalism enables us to determine the van der Waals contribution to the binding energy of the molecule to the interface.

\begin{figure}
\includegraphics[width=8cm]{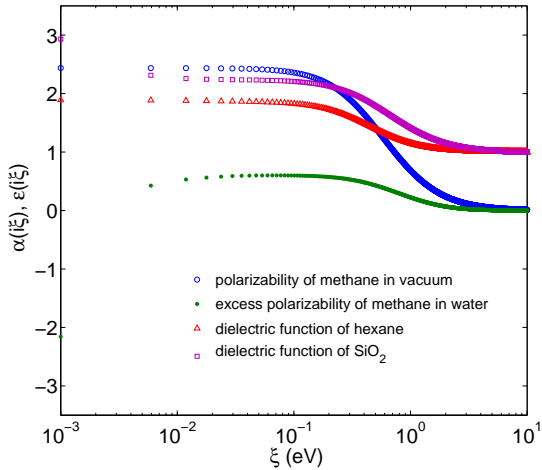}
\caption{(Color online) The polarizabilities of methane in units of \AA$^3$ in water and in vacuum. Note that the polarizability of methane is to a large degree isotropic and we show therefore only the mean average.  The polarizabilities corresponding to $n=0$ are shown on the $y$-axis. Also shown are the dielectric functions of hexane and SiO$_2$. }
\label{figu1}
\end{figure}

\section{Excess Polarizability of Molecule Solved in Water}

The excess polarizabilities at Matsubara frequencies  and  Gaussian radii for methane solved  in water were derived as in, for instance, papers by Parsons and Ninham\cite{ParsonsNinham2009,ParsonsNinham2010dynpol}. The polarizability of methane is to a high degree isotropic. Dynamic polarizabilities of the considered molecules in vacuum (see Fig.\,\ref{figu1}) were calculated
using \textsc{Molpro}\,\cite{MOLPRO2008} at a coupled cluster singles and
double (CCSD) level of theory. The excess polarisabilities, ${\alpha^*}(i{\xi})$, in water were obtained from the polarisabilities, ${\alpha}(i{\xi})$, in vacuum using the relation for a dielectric sphere embedded in a dielectric medium \cite{LandauLifshitz-ElectrodynContMedia-v8}, ${\alpha^*(i\xi_n)} = a^3 {({\varepsilon_a} - {\varepsilon_w})/({\varepsilon_a} + 2{\varepsilon_w})}$, where ${\varepsilon_w}$ is the dielectric function of water. ${\varepsilon_a}$ is the effective dielectric function of the molecular sphere, determined from the molecular polarisability in vacuum as  $ \varepsilon_a(i\xi_n) = 1 + 4\pi\alpha(i\xi_n) /V$,
where $V$ is the volume of the molecular sphere. The polarizabilities of methane in vacuum and excess polarizabilities of methane in water are shown in Fig. 1. The polarizabilities corresponding to $n=0$ are shown on the $y$-axis.  Due to rapid movements of the molecules in water we can use the orientation averaged excess polarizability for methane.

\begin{figure}
\includegraphics[width=8cm]{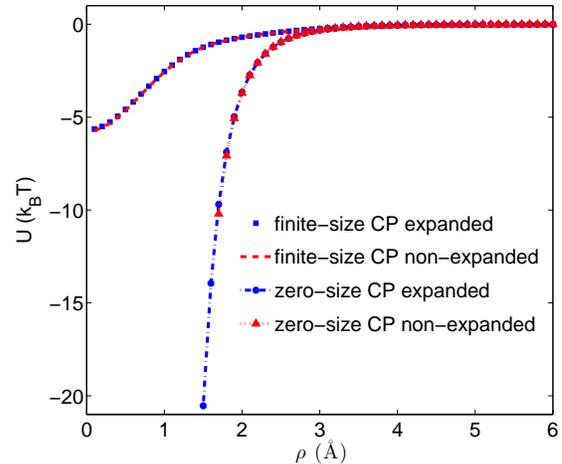}
\caption{(Color online) Comparing the series-expanded and the non-expanded theories of interaction of two methane molecules in water. The interation energy is in units of $k_BT$.}
\label{figu2}
\end{figure}

\section{The dielectric function of hexane, water, and SiO$_2$}
The dielectric function of hexane was calculated using a model dielectric function as given by T. Masuda \,\cite{T.Masuda}. 
The dielectric function of water was based on the extensive experimental data found in the {\it Handbook of Optical Constants of Solids II}\,\cite{Quer}. The dielectric function on the imaginary axis was obtained from the imaginary part of the function using the following version of the Kramers-Kronig dispersion relation
\begin{equation}
\varepsilon \left( {i\xi } \right) = 1 + \frac{2}{\pi }\int\limits_0^\infty  {d\omega \frac{{\omega {\varepsilon _2}\left( \omega  \right)}}{{{\omega ^2} + {\xi ^2}}}} .
\label{equ8}
\end{equation}
This relation is the result of the analytical properties of the dielectric function.\,\cite{Sernelius} In the integration we made a cubic spline interpolation of $\ln \left( {{\varepsilon _2}\left( \omega  \right)} \right)$ as a function of $\ln \left( \omega  \right)$.  

For $\rm SiO_2$ we calculated the dielectric function by means of first-principles atomistic models. 
Here the electronic structure of  $\rm SiO_2$ was calculated by employing a partial self-consistent GW method where  
the Green functions were updated iteratively while the screened Coulomb potential W was fixed.\,\cite{Kresse,Shishkin,Zhao} 
The electron-phonon coupling was initially neglected. 
From the electronic dispersion, the imaginary part of the dielectric functions was calculated from the linear response in the long wave length limit ($\lambda_q = 2\pi /q \to \infty$) through\,\cite{Gajdos}
 
\begin{equation}
\begin{array}{*{20}{l}}
{{\varepsilon _{2,j}}(\omega ) = \mathop {\lim }\limits_{q \to 0} \frac{{4{\pi ^2}{e^2}}}{{\Omega {q^2}}}\sum\limits_{c,v,{\bf{k}}} {2{w_{\bf{k}}}\delta ({E_c}({\bf{k}}) - {E_v}({\bf{k}}) - \hbar \omega )} }\\
{\quad \quad \quad \quad  \times \left\langle {{u_c}({\bf{k}} + {{\bf{e}}_j}q)\left| {{u_v}({\bf{k}})} \right.} \right\rangle {{\left\langle {{u_c}({\bf{k}} + {{\bf{e}}_j}q)\left| {{u_v}({\bf{k}})} \right.} \right\rangle }^*}}.
\end{array}
\end{equation}
Here, ${u_l}\left( {\bf{k}} \right)$ is the cell periodic part of the $l$:th wave function,  ${E_l}\left( {\bf{k}} \right)$ are the energies of the corresponding conduction ($l = c$) and valence ($l = v$) band states, and $\Omega $ is the volume of the primitive cell. The function ${w_{\bf k}}$ is the weight of the {\bf k}-points, and ${{\bf{e}}_j}$ is the unit vector in the cartesian coordinates. The dielectric function on the imaginary axis was obtained via the Kramers-Kronig dispersion relation, Eq.\,(\ref{equ8}), for each cartesian component.

The dielectric function in polar materials depends on the electron--optical-phonon coupling. 
We modeled this contribution employing the Lorentz model and the Kramers-Heisenberg formula\,\cite{Lorentz,Kuzmany}
 
\begin{equation}
{\varepsilon _{ph}}\left( \omega  \right) = 1 + \sum\limits_\eta  {\frac{{\left( {\omega _{{\rm{LO}},\eta }^2 - \omega _{{\rm{TO}},\eta }^2} \right){\varepsilon _\eta }}}{{\omega _{{\rm{TO}},\eta }^2 - {\omega ^2} - i{\gamma _\eta }\omega }}} ,
\end{equation}
using small phonon damping parameters ({\it i.e.}, ${\gamma _\eta } \to 0$). The parameters ${\omega _{{\rm{LO}},\eta }}$  and ${\omega _{{\rm{TO}},\eta }}$ are the phonon frequencies of the $\eta$:th longitudinal optical (LO) and transverse optical (TO) mode, respectively, and ${{\varepsilon _\eta }}$ is the high frequency dielectric constant of the TO phonon mode. 

The calculated dielectric functions of  $\rm SiO_2$ and hexane on the imaginary frequency axis are shown in Fig. 1. The static dielectric constant of SiO$_2$ is 3.9 which is in agreement with the measured data of 3.9-4.4.\,\cite{Bostr}

\section{Numerical Results: Casimir-Polder in solutions and Binding energy for molecules near planar interfaces}

As shown in Fig. 2, we find that two methane molecules in water attract each other with an estimated van der Waals binding energy around -0.23 $k_BT$ at room temperature. The value that we provide here corresponds to the separation distance ${\rho} = 2a$, {\it i.e.}, at the contact distance of the two molecules before their electron clouds start to overlap. This is deemed more physical.
Good agreement is found up to close separation distance between the series-expanded and the full non-expanded theories when finite size effects are accounted for. The presence of the background medium reduces the coupling between the two molecules, thereby making the series expansion a valid approximation.  In earlier work, we demonstrated that expanded and non-expanded theories give very different results for atom-atom interaction in vacuum. However, the coupling between two particles  in water is weaker than in vacuum, hence the series expanded theory is found to be a useful approximation for van der Waals interaction in water. This motivates us to use the series-expanded theory for interaction between a particle in a medium with a surface.
In Fig. 3, we compare the Casimir-Polder interactions of zero-sized and finite-sized methane molecule in water near a large SiO$_2$ surface. Incorporating the effects of finite molecule size in the calculation makes it possible to estimate the energy at contact distance, which is, in fact, the contribution of the van der Waals energy to the binding energy of the molecule to the surface.\,\cite{Maha}  In Fig. 4, we estimate the binding energy contribution of the van der Waals interaction of a methane molecule in water to different surfaces. Methane molecule in water is attracted to the SiO$_2$ surface while it is pushed away from the hexane and air surfaces. We provide estimates on the van der Waals contribution to the binding energy (at ${\rho} = a$) of methane molecules in water near different surfaces in Table I.

\begin{table}
\caption{The finite size van der Waals binding energy of methane molecules near different interfaces. All energies are in $k_BT$.}
\label{values}
\begin{tabular}{cccccc}
\tableline
\tableline
\multicolumn{1}{c}{Background  } &\multicolumn{1}{c}{Surface  } &\multicolumn{1}{c}{vdW  } \\
\tableline
Water& SiO$_2$ &-0.79 \\
Water& Hexane &0.14  \\
Water&Air &1.54 \\
\tableline
\tableline
\end{tabular}
\end{table}

\begin{figure}
\includegraphics[width=8cm]{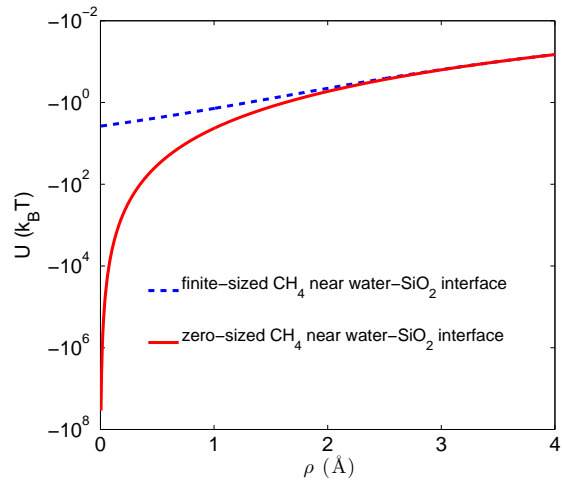}
\caption{(Color online) Casimir-Polder energy of methane in units of $k_BT$ near water-SiO$_2$ interface with and without finite-size effects. Note that including the effects of finite size removes the divergence at zero separation distance. }
\label{figu3}
\end{figure}

\begin{figure}
\includegraphics[width=8cm]{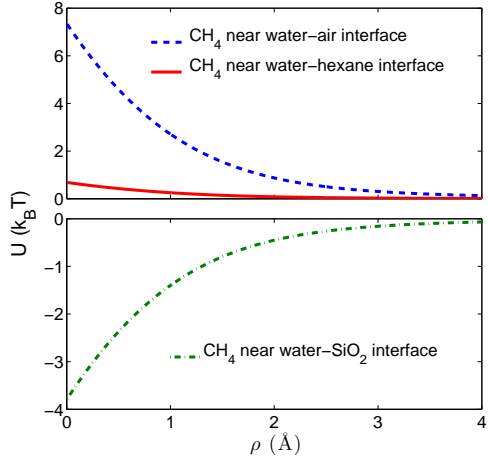}
\caption{(Color online) Comparing the Casimir-Polder energy of finite-sized methane in water near different surfaces. SiO$_2$ surface attracts the methane molecule while hexane and air surfaces repulse it. }
\label{figu4}
\end{figure}

\section{Conclusions}

We have discussed the finite size dependent Casimir-Polder interaction between polarizable particles and one such particle and a planar surface. This enables us to estimate contributions from dispersion interactions to the binding energy. These interactions provide a mechanism for selective adsorption or desorption of different molecules, near nanosurfaces and semi-planar surfaces solved in liquids. Our result demonstrates that the expanded theory often works surprisingly well as an approximation for a proper non-expanded theories; at least in the particular case of two finite sized methane molecules in water. One shall stress that it is only when finite size is accounted for that the interaction remains finite at close separation distance and the expanded theory provides a good estimate for all separations.

\section{Acknowledgements}
P.T. gratefully acknowledges support from the European Commission; this publication reflects the views only of the authors, and the Commission cannot be held responsible for any use which may be made of the information contained therein. M.B., C.P., and A.M.-S. acknowledge support from the Research Council of Norway (Project: 221469). C.P. and M.B. acknowledge support from VR (Contract No. 90499401). MB also thanks the Department of Energy and Process Engineering (NTNU, Norway) for financial support. We acknowledge access to high-performance computing resources via SNIC/SNAC and NOTUR. Ab initio calculations were undertaken at the NCI National Facility in Canberra, Australia, which is supported by the Australian Commonwealth Government.

\begin{figure}
\includegraphics[width=8cm]{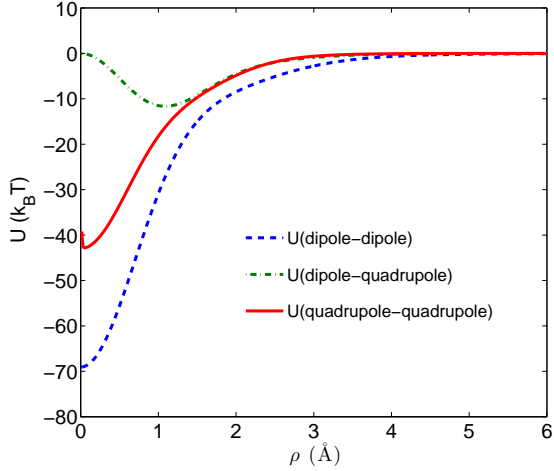}
\caption{(Color online) Dipole-dipole, dipole-quadrupole and quadrupole-quadrupole contributions to the interaction between two methane molecules in vacuum. }
\label{figu5}
\end{figure}

\appendix
\section{Derivation of the non-retarded Green's functions with finite size effects}

We briefly show how the non-retarded Green's functions are derived. The Green's function obtained from solving the inhomogeneous Helmholtz equation for a coupled system of two neutral particles at positions ${{\bf R}_1}$ and ${{\bf R}_2}$ is \,\cite{Maha, Buhmann12a}





\begin{equation}
\begin{array}{l}
{\bf{T}}\left( {{{\bf{R}}_1},{{\bf{R}}_2};i\xi } \right)\\
 = \frac{1}{{{{\left( {2\pi } \right)}^3}}}\smallint \frac{{{d^3}k[({\xi ^2}/{c^2})\tilde I + {\bf{k}} \cdot {\bf{k}}]}}{{[({\xi ^2}/{c^2}) + {k^2}]}}\\
 \times {e^{i{\bf{k}} \cdot ({{\bf{R}}_1} - {{\bf{R}}_2})}} \times \int {{e^{ - i{\bf{k}} \cdot {{\bf{R}}_3}}}\pmb \alpha ({{\bf{R}}_3},\xi ){d^3}{R_3}} 
\end{array}
\end{equation}
where $ {{\pmb {\alpha}}({\bf R}, {\xi})} = {{\tilde I}{\alpha}({\xi})f(R)} $. As mentioned before in Sec. II, ${\alpha}({\xi})$ is the polarizability at the Matsubara frequencies ${\xi}$ and, the function $f(R)$ is assumed to be a gaussian given by
\begin{equation}
f(R) = \frac{1}{{\pi}^{3/2}{a^3}} {e^{-{R^2}/{a^2}}},
\end{equation}
where $a$ is the gaussian radius of the particle. 

In the non-retarded limit ($c \to \infty$), we obtain using the above equations and the spherical coordinates

\begin{equation}
\begin{array}{l}
{ T_{xx}}\left( {{{\bf{R}}_1},{{\bf{R}}_2} } \right) = { T_{yy}}\left( {{{\bf{R}}_1},{{\bf{R}}_2} } \right)\\
 = \frac{{1}}{{{{\left( {2\pi } \right)}^3}}}\int {{k^2}\sin \theta d\theta d\varphi dk\frac{{ - k_x^2}}{{ - {k^2}}}{e^{ikR\cos \theta }}{e^{ - {k^2}{a^2}/4}}} \\
 = \frac{{1}}{{{{\left( {2\pi } \right)}^3}}}\int {\sin \theta d\theta d\varphi dkk_x^2{e^{ikR\cos \theta }}{e^{ - {k^2}{a^2}/4}}}  = \\
 = \frac{{1}}{{{{\left( {2\pi } \right)}^3}}}\int {\sin \theta d\theta d\varphi dk{k^2}{{\sin }^2}\theta {{\cos }^2}\varphi {e^{ikR\cos \theta }}{e^{ - {k^2}{a^2}/4}}} \\
 = \left| \begin{array}{l}
\cos \theta  = x\\
dx =  - \sin \theta d\theta
\end{array} \right| = \\
 = \frac{{1}}{{{{\left( {2\pi } \right)}^2}}}\frac{1}{2}\int\limits_{ - 1}^1 {dx\int\limits_0^\infty  {dk{k^2}\left( {1 - {x^2}} \right){e^{ikRx}}{e^{ - {k^2}{a^2}/4}}} } \\
 = \frac{{1}}{{{{\left( {2\pi } \right)}^2}}}\int\limits_0^1 {dx\int\limits_0^\infty  {dk{k^2}\left( {1 - {x^2}} \right)\cos \left( {kRx} \right){e^{ - {k^2}{a^2}/4}}} } \\
 = \frac{{1}}{{{{\left( {2\pi } \right)}^2}}}\frac{{4\sqrt \pi  }}{{{a^3}}}\int\limits_0^1 {dx\left( {1 - {x^2}} \right)\left[ {\frac{1}{2} - {{\left( {\frac{R}{a}x} \right)}^2}} \right]\exp \left[ { - {{\left( {\frac{R}{a}x} \right)}^2}} \right]} \\
 = \frac{{1}}{{{{\left( {2\pi } \right)}^2}}}\frac{{4\sqrt \pi  }}{{{a^3}}}\left[ {\frac{1}{4}\frac{{\sqrt \pi  {\mathop{\rm erf}\nolimits} \left( {\frac{R}{a}} \right) - 2\left( {\frac{R}{a}} \right){e^{ - {{\left( {\frac{R}{a}} \right)}^2}}}}}{{{{\left( {\frac{R}{a}} \right)}^3}}}} \right]\\
 = \frac{{1}}{{{{\left( {2\pi } \right)}^2}}}\frac{{\sqrt \pi  }}{{{R^3}}}\left[ {\sqrt \pi  {\mathop{\rm erf}\nolimits} \left( {\frac{R}{a}} \right) - 2\left( {\frac{R}{a}} \right){e^{ - {{\left( {\frac{R}{a}} \right)}^2}}}} \right],
\end{array}
\end{equation}
\begin{equation}
\begin{array}{l}
{ T_{zz}}\left( {{{\bf{R}}_1},{{\bf{R}}_2} } \right)\\
 = \frac{{1}}{{{{\left( {2\pi } \right)}^3}}}\int {{k^2}\sin \theta d\theta d\varphi dk\frac{{ - k_z^2}}{{ - {k^2}}}{e^{ikR\cos \theta }}{e^{ - {k^2}{a^2}/4}}} \\
 = \frac{{1}}{{{{\left( {2\pi } \right)}^3}}}\int {\sin \theta d\theta d\varphi dkk_z^2{e^{ikR\cos \theta }}{e^{ - {k^2}{a^2}/4}}}  = \\
 = \frac{{1}}{{{{\left( {2\pi } \right)}^3}}}\int {\sin \theta d\theta d\varphi dk{k^2}{{\cos }^2}\theta {e^{ikR\cos \theta }}{e^{ - {k^2}{a^2}/4}}} \\
 = \left| \begin{array}{l}
\cos \theta  = x\\
dx =  - \sin \theta d\theta
\end{array} \right| = \\
 = \frac{{1}}{{{{\left( {2\pi } \right)}^2}}}\int\limits_{ - 1}^1 {dx\int\limits_0^\infty  {dk{k^2}{x^2}{e^{ikRx}}{e^{ - {k^2}{a^2}/4}}} } \\
 = \frac{{1}}{{{{\left( {2\pi } \right)}^2}}}2\int\limits_0^1 {dx\int\limits_0^\infty  {dk{k^2}{x^2}\cos \left( {kRx} \right){e^{ - {k^2}{a^2}/4}}} } \\
 = \frac{{1}}{{{{\left( {2\pi } \right)}^2}}}\frac{{8\sqrt \pi  }}{{{a^3}}}\int\limits_0^1 {dx{x^2}\left[ {\frac{1}{2} - {{\left( {\frac{R}{a}x} \right)}^2}} \right]\exp \left[ { - {{\left( {\frac{R}{a}x} \right)}^2}} \right]} \\
 = \frac{{1}}{{{{\left( {2\pi } \right)}^2}}}\frac{{8\sqrt \pi  }}{{{a^3}}}\left[ {\frac{1}{4}\frac{{ - \sqrt \pi  {\mathop{\rm erf}\nolimits} \left( {\frac{R}{a}} \right) + 2\left( {\frac{R}{a}} \right){e^{ - {{\left( {\frac{R}{a}} \right)}^2}}} + 2{{\left( {\frac{R}{a}} \right)}^3}{e^{ - {{\left( {\frac{R}{a}} \right)}^2}}}}}{{{{\left( {\frac{R}{a}} \right)}^3}}}} \right]\\
 = \frac{{1}}{{{{\left( {2\pi } \right)}^2}}}\frac{{2\sqrt \pi  }}{{{R^3}}}\left[ { - \sqrt \pi  {\mathop{\rm erf}\nolimits} \left( {\frac{R}{a}} \right) + 2\left( {\frac{R}{a}} \right)\left[ {1 + {{\left( {\frac{R}{a}} \right)}^2}} \right]{e^{ - {{\left( {\frac{R}{a}} \right)}^2}}}} \right],
\end{array}
\end{equation}
where $R=|{\bf R}_1-{\bf R}_2|$. To make it consistent with the definitions in the main text, we have also normalized the Green's functions by removing the factor $\alpha (i\xi)$. 

\section{Multipolar Dispersion Interaction between Methane Molecules in Vacuum}

At the zero separation limit, in addition to the dipole-dipole interaction, effects due to multipole interactions and wavefunction overlap begin to contribute to the total interaction energy. In Fig. 5, we show the various contributions due to the dipole-dipole, the dipole-quadrupole and the quadrupole-quadrupole interactions between two methane molecules in vacuum using expanded theory in the non-retarded limit. \,\cite{Richardson} (The dipole-dipole interaction is a factor of 2 too small in Ref. \cite{Richardson} and the prefactor in their Eq.\,(34) should be $16/9\pi$ instead of $8/4\pi$). As can be observed from the figure, major contribution comes from the dipole-dipole interaction at all separations.  A recent paper by DiStasio {\it et al.} considers orbital overlap explicitly using the density functional (quantum mechanical) theory of interaction potential between finite-sized quantum harmonic oscillators. \,\cite{DiStasio} Their result agrees with ours in that the finite molecule size renders the interaction finite at zero molecule-molecule separation.

\end{document}